A brief history of the Fail Safe Number in Applied Research

Moritz Heene

University of Graz, Austria

*Introduction*

Rosenthal's (1979) Fail-Safe-Number (FSN) is probably one of the best known statistics in the context of meta-analysis aimed to estimate the number of unpublished studies in meta-analyses required to bring the meta-analytic mean effect size down to a statistically insignificant level. Already before Scargle's (2000) and Schonemann & Scargle's (2008) fundamental critique on the claimed stability of the basic rationale of the FSN approach, objections focusing on the basic assumption of the FSN which treats the number of studies as unbiased with averaging null were expressed throughout the history of the FSN by different authors (Elashoff, 1978; Iyengar & Greenhouse, 1988a; 1988b; see also Scargle, 2000). In particular, Elashoff's objection appears to be important because it was the very first critique pointing directly to the central problem of the FSN: "R & R claim that the number of studies hidden in the drawers would have to be 65,000 to achieve a mean effect size of zero when combined with the 345 studies reviewed here. But surely, if we allowed the hidden studies to be *negative*, on the average no more than 345 hidden studies would be necessary to obtain a zero mean effect size" (p. 392). Thus, users of meta-analysis could have been aware right from the beginning that something was wrong with the statistical reasoning of the FSN. In particular, from an applied research perspective, it is therefore of interest whether any of the fundamental objections on the FSN are reflected in standard handbooks on meta-analysis as well as -and of course even more importantly- in meta-analytic studies itself.

The aim of the present commentary is therefore twofold. Firstly, to review the discussion of the FSN in standard handbooks on meta-analysis. Secondly, to summarize results from a trend analysis, ranging from 1979 until 2008, on the number of meta-analyses from various research fields which used the FSN.

*Discussion of the FSN in selected handbooks on meta-analysis*

Based on an analysis of citation frequency, the four most cited handbooks were selected for this overview: a) Hedges & Olkin (1985), b) Hunter & Schmidt (1990, 2004) and c) Cooper & Hedges (1994). In the following section, book sections explicitly related to the FSN are reported and brief commentaries given.

*Hedges & Olkin (1985)*

After introducing the reader into the underlying logic of the FSN, Hedges & Olkin (1985, p. 306) discussed the FSN as follows: "The weakness in this argument is its reliance on the combined test statistic. We argued in Chapter 3 that combined significance tests seldom answer the question of interest to research reviewers. Rejection of the combined null hypothesis implies at least one study has nonzero effect. Because this alternative hypothesis is usually not substantively meaningful, methods that increase the plausibility of the hypothesis are not particularly useful".

Thus, Hedges & Olkin's critique regards the null hypothesis ritual of setting up and testing a trivial null hypothesis of zero effect (which one is pretty sure is wrong) as most damaging to the application of the FSN. The more fundamental assumption of the FSN of additional unlocated studies averaging $z = 0$ needed to bring down the observed test statistic to an insignificant level is not addressed.

*Hunter & Schmidt (1990, 2004)*

In their 1st edition of their book on meta-analysis Hunter & Schmidt (1990, p. 512) write: "If in this example there were originally 100 studies ($k = 100$) instead of 10, then 6,635! Over 6,000 studies would be required to raise the combined *p*-value to .05. In most research areas, it is inconceivable that there could be over 6,000 "lost" studies. Examples of use of this method in personnel selection research are given in Callender and Osburn (1981) and Schmidt, Hunter, and Caplan (1981b)"

In their 2nd edition (2004, p. 448) they add the following critical remark on the FSN to the same paragraph: "This number [i.e., the fail-safe-number] typically turns out to be very large …. However, the file drawer technique … is a fixed-effects model and, therefore, yields accurate results only if the underlying correlation (or *d* value) is identical in all studies. If population values of $\rho$ and $\delta$ vary across studies, … the number of studies needed to make the combined *p*-value just barely significant is much smaller than the number provided by the file drawer analysis". Hence, Hunter and Schmidt's main critique points to the implicit assumption made by the FSN that all analyzed studies as a whole are considered to have been conducted under similar conditions with similar subjects and differ only in regard to their statistical power to detect the effect of interest. The central assumption of the FSN of additional unlocated studies averaging $z = 0$ is not addressed.

*Cooper & Hedges (1994)*

Begg (1994, p. 406), writes: "The advantage of the file-drawer method is that it is simple and easily interpretable. A disadvantage is the assumption that the results of the missing studies are centered on the null hypothesis, an assumption that seems artificial and is basically a device to permit an interpretable correction. Also, the method is not influenced in any way by the evidence in favor of bias in the data. That is, is not influenced by the shape of the funnel graph, for example. The correction is entirely based on the conclusiveness of the *p* value and the number of component studies in the meta-analysis, relative to the number that might be missing".

Thus, in accordance with Hedges & Olkin (1985), Begg's critique focuses mainly on the null hypothesis ritual of testing a trivial or at least "artificial", null hypothesis.

Summarizing the citations above it is obvious that all these authors did not address the elementary fact that, using the example of Rosenthal & Rubin, if 345 is 5% of all studies, then the total number of studies must have been $345/.05 = 6,900$, not 65,000.

*Results from a trend analysis on the use of the FSN in meta-analyses*

It may be asked whether the existent critique on the FSN as mentioned in the introduction had any impact on the use of the FSN in applied research. To answer this question a literature research (based on published meta-analytic results from 1979-2008) was conducted via the *Web of Science Database* resulting in 520 meta-analyses from various research fields having used the FSN. These meta-analytical studies were then classified in five categories: Psychology, Medicine, Health Sciences, Psychiatry and "Others" (the latter category included research field such as Criminology, Political Sciences and Pharmacology, Epidemiology, Veterinary Medicine and Consumer Research etc.)[1]. Periodicity of the FSN was counted in four-year intervals and the results were cross-tabulated with the five research fields mentioned above. Figure 1 shows the periodicity of the FSN over the last three decades, separated by research field.

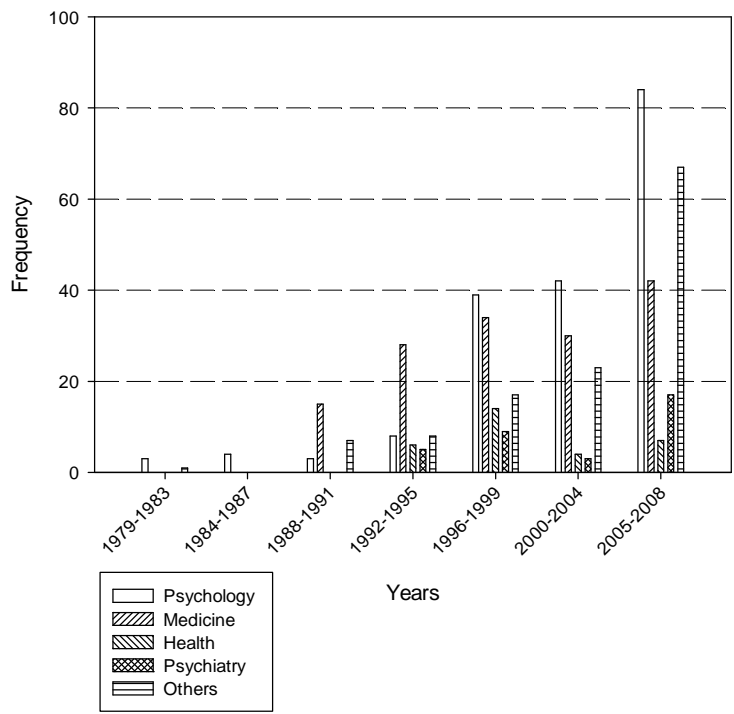

*Figure 1: Periodicity of the FSN in different research areas from 1979 to 2008*

---

[1]: The complete list of the meta-analyses used in this study can be obtained from the author.

Aggregating the results over all research fields and conducting a trend analysis with a linear and an exponential regression yielded an excellent fit for the exponential model (*MNSQ-Residual*$_{exponential\ Regression}$ = .07, $R^2$ = .87, $F(1, 518)$ = 3454.30, $p < .001$; *MNSQ-Residual*$_{linear\ Regression}$ = 708.24, $R^2$ = .84, $F(1, 518)$ = 2810.50, $p < .001$).

Given these results, it is quite obvious that the detection of the faulty FSN method and the derived conclusions about the instability of meta-analytic results even in the presence of a modest number of unpublished studies (Scargle, 2000; Schonemann & Scargle, 2008) went largely unheeded. More importantly: the exponential increase of the popularity of the FSN since the appearance of numerous handbooks shows that the FSN is still regarded as a useful tool for putting one's meta-analytic result on a sound basis.

References

Begg, C. B. (1994). Publication Bias. In H. Cooper & L. V. Hedges (Eds.), *The handbook of research synthesis*. New York: Russell Sage Foundation.

Elashoff, J. D. (1978). Commentary. *The Behavioral and Brain Sciences, 1*, 392.

Hedges, D. W., & Olkin, I. (1985). *Statistical Methods for Meta-Analysis*. Orlando: Academic Press.

Hunter, J. E., & Schmidt, F. L. (1990). *Methods of meta-analysis: Correcting Error and Bias in Research Finding* (1st ed.). Newbury: Sage.

Hunter, J. E., & Schmidt, F. L. (2004). *Methods of meta-analysis: Correcting Error and Bias in Research Finding* (2nd ed.).

Iyengar, S., & Greenhouse, J. B. (1988a). Rejoinder. *Statistical Science, 3*, 133-135.

Iyengar, S., & Greenhouse, J. B. (1988b). Selection models and the file-drawer problem. *Statistical Science, 3*, 109-135.

Rosenthal, R. (1979). *The file drawer problem and tolerance for null results*: Psychological Bulletin Vol 86(3) May 1979, 638-641.

Scargle, J. D. (2000). Publication bias: The "File Drawer" problem in scientific inference. *Journal of Scientific Exploration, 14*, 91-106.

Schonemann, P. H., & Scargle, J. D. (2008). A Generalized Publication Bias Model. *Chinese Journal of Psychology, 50*(1), 21-29.